\documentclass[aps,pra,twocolumn,groupedaddress,showpacs,showkeys,superscriptaddress]{revtex4-1}

\usepackage{amsmath}
\usepackage{amssymb}
\usepackage{graphicx}
\usepackage{epstopdf}

\usepackage{subcaption}
\usepackage{mwe}

\usepackage{media9}
\usepackage{hyperref}

\begin{document}

\title{Nonlinear optical switching in hybrid plasmonic waveguides}

\author{Ivan A. Pshenichnyuk}
\email[correspondence address: ]{i.pshenichnyuk@skoltech.ru}
\author{Fahmy Yousry}
\author{Daniil S. Zemtsov}
\author{Sergey S. Kosolobov}
\author{Vladimir P. Drachev}
\affiliation{Skolkovo Institute of Science and Technology, Moscow 121205, Russian Federation}

\date{\today}

\begin{abstract}
We investigate an optical switching mechanism for applications in active integrated photonic circuits. The mechanism utilizes a large nonlinearity of indium-tin-oxide (ITO) in epsilon-near-zero (ENZ) regime. The effect of optically induced switching is investigated in hybrid plasmonic waveguides (HPWG) with ITO layer that are often used as a platform for various active photonic components, including optical modulators. To study the effect, nonlinear Maxwell equations are solved numerically in time and frequency domains. The all-optical mechanism of switching in HPWG is quantitatively compared with an electrical gating that is often used to manipulate hybrid modes. It is shown that the optical pumping could potentially cause more efficient switching. The influence of light intensity on HPWG properties and interplay between hybrid modes are investigated numerically in different regimes. Transmittance and phase variation are calculated as a function of light intensity for Si and Si$_3$N$_4$ waveguides. It is shown that the transmittance and its intensity dependence are strongly affected by the ITO thickness. A significant for applications step-like variation of phase and transmittance with intensity is detected. The possibility to control the excitation conditions for surface plasmon polaritons (SPP) in HPWG using active layer of ITO in nonlinear regime is demonstrated and discussed.
\end{abstract}

\maketitle

\section{Introduction} \label{sec_intro}

Successful development of integrated photonic circuits requires compact and efficient active elements, such as modulators, tunable ring resonators, phase shifters etc. There are numerous electro-optical \cite{liu-2015} and all-optical \cite{chai-2017} switching mechanisms discussed in the literature. Proposed devices compete for switching speed, energy consumption, optical losses, and dimensions. The general challenge in the domain of all-optical switching devices is in relatively small nonlinearities of traditional materials that results in large sizes and high intensity pumping required for switching. At the same time, novel materials with sufficiently large nonlinearities and convenient for fabrication are discovered. It was demonstrated that nonlinearities can be significantly amplified in epsilon-near-zero (ENZ) regime \cite{reshef-2019}. Therefore, ENZ materials and their peculiar properties (both linear and nonlinear) are considered as exceptionally promising in modern photonics \cite{wu-2021}. Here we study in comparison all-optical switching and electro-optical switching in the same hybrid plasmonic waveguide structure and report promising peculiarities for the all-optical case.

Transparent conductive oxides (TCO), and indium-tin-oxide (ITO) in particular, are known for a long time as promising materials for electro-optical circuits \cite{jaffray-2022}. Intermediate concentration of the charge carriers makes them transparent at optical range of wavelengths and electrically conductive at the same time. Depending on applications demands, concentration can be varied using doping. Many ITO-based active devices presented in the literature  \cite{melikyan-2011, sorger-2012, pshenichnyuk-2019b}, rely on the electronic activation mechanism. An external electric field is applied to induce an accumulation layer at ITO/insulator boundary. That affects a refractive index and can be used to vary an optical response. Significant changes of the optical properties may take place in the vicinity of epsilon-near-zero (ENZ) point, where the real part of the permittivity crosses zero and changes the material response from dielectric to metallic.

ENZ switching behavior can be enhanced using plasmonics. In particular, the concept of a hybrid plasmonic waveguide (HPWG) has been proved to be useful for applications \cite{alam-2014, alam-2013,pshenichnyuk-2018c}. An additional layer of plasmonic material (like silver or gold) deposited on top of ENZ sandwich significantly increases the interaction of light with an accumulation layer. It also allows to make devices smaller and faster. The concept is widely used for electro-optical modulation \cite{sorger-2012, pshenichnyuk-2019,zemtsov-2023}.

Accumulation layers in TCO are usually quite thin. Their structure is known from both classical and quantum models \cite{sinatkas-2017,gao-2018b}. Specifically, the layer thickness to be of the order of few nanometers. Practically it means that only a small fraction of ITO is active at the applied voltage. For a $10$ nm thick layer of ITO only about 10$\%$ participate in switching. Since the local variation of the refractive index can be quite large for both real and imaginary parts this is enough for a certain class of applications to provide phase and amplitude switching. This is especially true for HPWG geometries that amplify the effect. On the other hand, the induced integral changes in optical properties are not sufficient to cause significant qualitative changes in a set of available modes. For example, usually it is not enough to make certain modes forbidden (except some exotic modes, like thin film plasmonic pairs supported by the accumulation layer \cite{pshenichnyuk-2021,puplauskis-2021}). Thus, it is desirable to find a stronger activation mechanism that may cause refractive index variation in a larger fraction of volume.

A larger fraction can be controlled with the all-optical activation mechanism. It was shown that TCOs possess strong nonlinear properties in ENZ regime \cite{alam-2016,caspani-2016}. According to the experimental data, the real (imaginary) part of ITOs refractive index variation may reach the value $0.7$ ($0.15$), while an intensity is varied between $0$ and $250$ GW/cm$^2$. Taking a linear approximation at small intensities one may evaluate the $n_2$ nonlinear coefficient to be of the order of $0.01$ cm$^2$/GW. It is much stronger than the usual out-of-ENZ Kerr coefficient for ITO (${\sim}5\times10^{-5}$ cm$^2$/GW ) and corresponding coefficient for silicon waveguides \cite{tsang-2008,timurdogan-2017}. Such changes in the refractive index take place, presumably, in the whole bulk of ITO reached by an electro-magnetic field. It makes optical activation potentially attractive to realize strong switching. Another experimental fact, promising for applications is that the ENZ nonlinearity in ITO is extremely fast \cite{alam-2016}. Finalizing the motivation list it is essential to say, that entering a nonlinear domain one may expect to discover new interesting physics. One bright example is related to room temperature polariton condensates \cite{lagoudakis}, where a nonlinear wave equation (Gross-Pitaevskii equation) successfully predicts a new class on intriguing solutions like solitons and quantum vortices  \cite{pshenichnyuk-2017, pshenichnyuk-2015}.

ENZ nonlinearity in ITO is probed experimentally and already applied in the field of metasurfaces \cite{alam-2018,guo-2016a}. For instance, generation of higher harmonics from ENZ materials is studied \cite{yang-2019}. Plasmonic enhancement of the nonlinear behavior using Au meta-atoms on ITO substrate is reported \cite{deng-2020}. Applications in the field of integrated photonics are on the way, but the existing works are mainly theoretical. TCO based absorption modulator is introduced by \citet{li-2021a}. In the suggested model ENZ layer is first created using electrical gating and then it is optically pumped (both activation mechanisms discussed above are involved). It allows the modulator to operate at different wavelengths (that correspond to different ENZ densities). At the same time, as in the case of plasmonic electro-optical modulators, only a small fraction of ITO is available for switching. Optically tuned nonlinear phase shifter is suggested by \citet{navarroarenas-2022}. A layer of active material is placed on top of Si waveguide with additional insulating SiO$_2$ layer between them. Even without plasmonic amplification, $\pi$ phase shift is achieved for just $6$ $\mu$m long device. An active directional coupler, where the switching is performed using ENZ TCO layer is reported in the work of \citet{sha-2022}. Finally, we would like to add that the idea to place a layer of active material on top of a waveguide to maintain switching emerges also in parallel fields of optics. In non-integral photonics, for instance, a layer of carbon nanotubes with a large nonlinear coefficient placed on top of an optical fiber allows to achieve a nontrivial intensity dependance of transmittance \cite{davletkhanov-2023}. 

High mobility CdO is often assumed as an active material in theoretical works. The obvious advantage is that it introduces small optical losses. On the other hand, it is sufficiently toxic \cite{hossain-2012,sreekanth-2016}. The analysis of various TCO performance in the context of switching is available in literature \cite{navarroarenas-2023}. In our work we rely on ITO since it is easily available photonic material with a well developed workflow. Nonlinearity in ITO is also well studied experimentally and the results of measurements are available in literature \cite{alam-2016}. Such a choice of an active material allows implementing a performance comparison with existing ITO based plasmonic electro-optical modulators \cite{zemtsov-2023}.

\begin{figure}
\centerline{\includegraphics[width=0.50\textwidth]{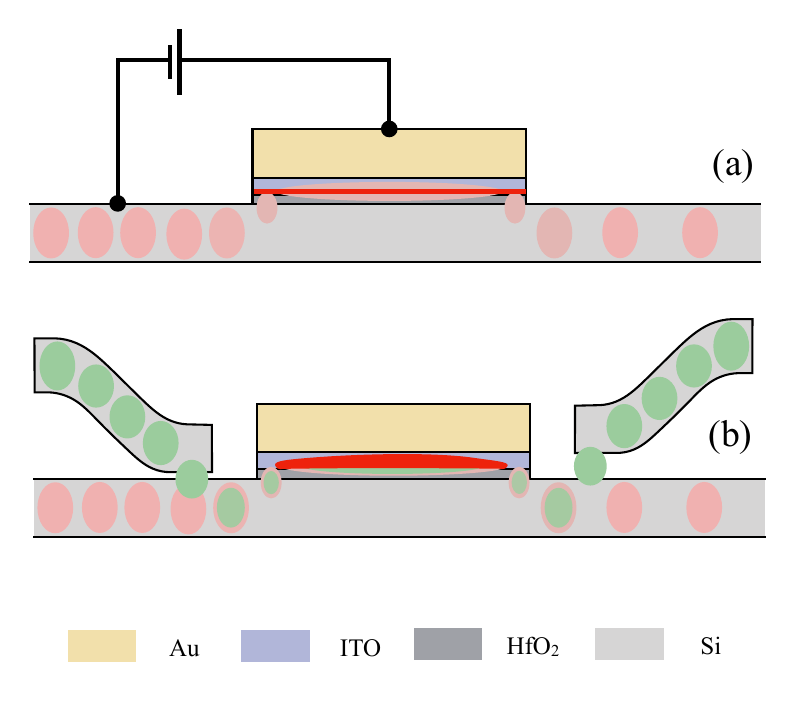}}
\caption {Schematic representation of electrical (a) and optical (b) activation mechanisms in HPWG based modulator.
\label{fig1_scheme}}
\end{figure}

\begin{figure*}
\centerline{\includegraphics[width=1.0\textwidth]{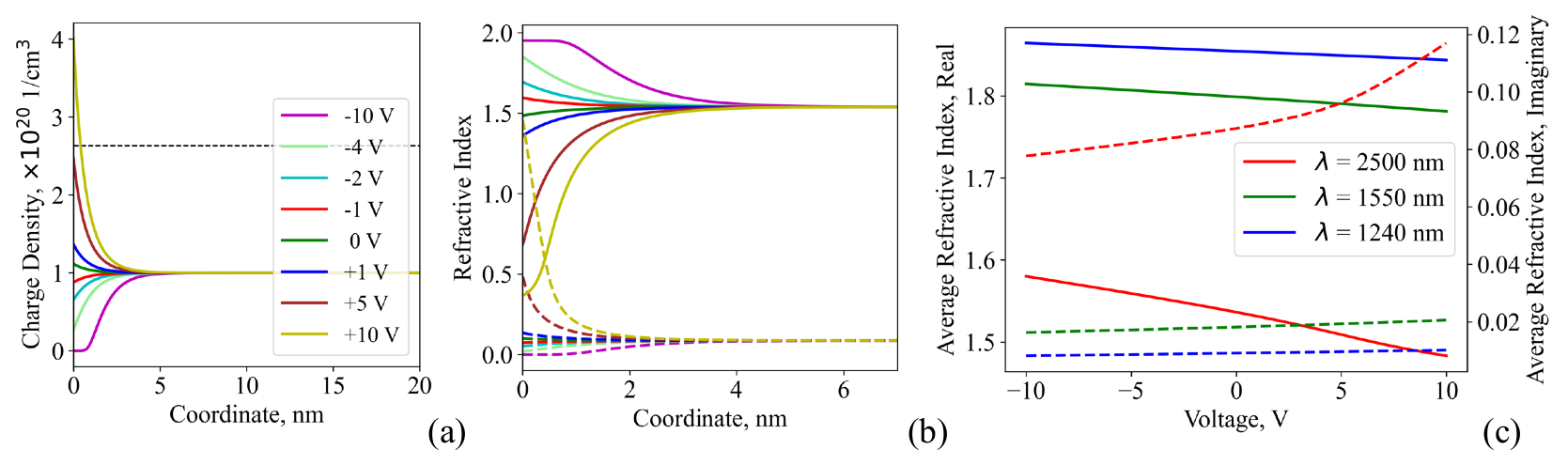}}
\caption {Electron density distribution in the layer of ITO (a) and corresponding variation of the refractive index real  and imaginary parts (b) at different voltages. Average refractive index variation in ITO as a function of voltage (c) for different wavelengths allows to evaluate the electrical switching strength. Solid and dashed color lines correspond to real and imaginary parts respectively. Horizontal black dashed line in (a) shows ENZ concentration.
\label{fig2_dd}}
\end{figure*}

In this paper plasmon-amplified nonlinear switching effects in ITO based HPWG is investigated. We start from the quantitative comparison between the electrical and all-optical switching schemes and show that the latter one is promising since it allows to use a larger fraction of an active material. Drift-diffusion equations are solved to model the formation of an accumulation layer at ITO/insulator boundary. Three different computational approaches based on Maxwell equations are used for the analysis of optical switching. Next, we evaluate the possibility to use nonlinear HPWGs as phase shifters and amplitude modulators. To perform the analysis we introduce a set of models. Typical HPWG sandwich composition includes a waveguide covered by a layered structure containing insulator, ITO, and Au. Waveguides of different geometry made from Si and Si$_3$N$_4$, as well as variable ITO layer thickness are considered. For the selected set of models we compute transmittance and phase as a function of intensity. Best results for amplitude and phase variation are obtained in different geometries. A significant for switching applications sudden variation of characteristics at a certain value of intensity is detected. The emphasis in the analysis is made on a strong switching, when some of the available modes can be attenuated. In particular, the possibility to prohibit SPP excitation at certain range of intensities is discussed. It is discovered that the thickness of ITO plays a significant role in this context, since appropriately chosen ENZ layer can be used to 'screen' the metallic surface and control the excitation of plasmons. The description of models and computation methods are collected in Sec.~\ref{sec_models} of the manuscript, while the results of our calculations are presented in Sec.~\ref{sec_results}.

\section{Models and Methods} \label{sec_models}

In properly tuned HPWG an ordinary waveguide mode can be transformed into a surface plasmon and back. The plasmonic state is characterized by subwavelength confinement and amplification of light. In this form it can be efficiently manipulated using a relatively thin layer of active material. In the electrical switching mechanism, often used for plasmonic modulators \cite{pshenichnyuk-2019b}, the active layer changes its properties under the influence of the field effect. It is schematically depicted in Fig.~\ref{fig1_scheme}a. A thin accumulation layer (red area) with an altered refractive index (both real and imaginary parts) is formed under the applied voltage at ITO/insulator boundary.

In the optical switching mechanism (Fig.~\ref{fig1_scheme}b) a variation of the refractive index takes place under the influence of a high intensity pump pulse. An average refractive index variation in this case depends on the intensity distribution in ITO (red area in Fig.~\ref{fig1_scheme}b). Since there is no strict thickness limitation for such a mechanism, it can potentially provide more efficient switching. Moreover, the local field intensity inside the sandwich is enhanced by the plasmonic effect that improves the switching contrast. A high intensity switching pulse may be injected using an additional directionally coupled waveguide (as depicted in Fig.~\ref{fig1_scheme}b, green color) or, alternatively, it may be transmitted using the same waveguide where the signal propagates. One may think about using slightly different wavelength for the pump, or, with certain modifications of the scheme, different polarization and mode number. In the calculations below for the simplicity of the analysis we consider only the pump signal that propagates through the 'main' waveguide with no additional geometry modifications. 
Or, in other words, we consider a self modulation of the pump pulse. The insulating layer inside the sandwich is needed to create an electric field in the case of electrical mechanism. It also can influence the coupling between plasmonic and waveguide modes in both optical and electrical cases. Thus, we use it to maintain a consistency between two schemes for comparison reasons.

In our calculations we consider HPWGs with various geometrical and material parameters. For the analysis in this paper we present a set of models with a waveguides thickness $230$, $400$ and $500$ nm and ITO layer thickness $20$, $30$ and $40$ nm. It is representative for the discussion of nonlinear switching including both phase and amplitude effects (they manifest themselves in different regimes). At the same time, numbers are optimized to simplify possible subsequent experimental fabrication and tests. Slab waveguides are considered everywhere in this paper. They are easy to implement experimentally and analyze theoretically since the confinement in the horizontal plane of the chip does not play a role. To make computations for slab waveguides it is enough to consider 2D models. Practically it means that the waveguides in the experiment should be sufficiently wide ($10$ $\mu$m and more, according to our experimental check). We also compared different materials for a waveguide. Two popular cases in integrated photonics, namely Si and Si$_3$N$_4$, are considered here. The latter one is more transparent at shorter wavelengths (compared with standard $1550$ nm). It is also more 'surface plasmon friendly' (see the discussion in the next section).

\begin{figure*}
\centerline{\includegraphics[width=1.0\textwidth]{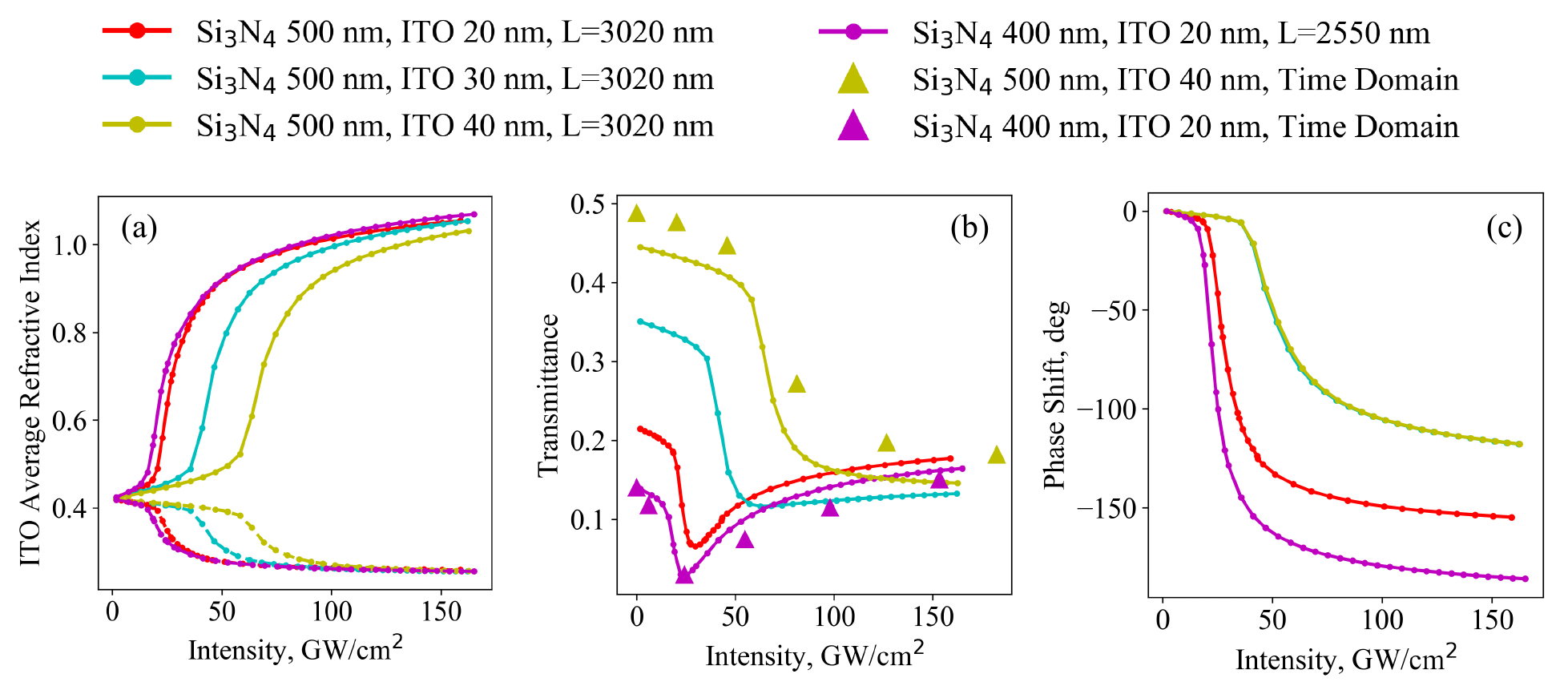}}
\caption {Average refractive index variation in ITO caused by optical pumping with different intensities (a) allows to evaluate optical switching strength quantitatively. Results are presented for silicon nitride based HPWGs in different geometries. Transmittance (b) and phase variation (c) in corresponding HPWG structures are plotted as a function of intensity. Solid lines correspond to frequency domain calculations, while triangular markers denote time domain calculation results.
\label{fig3_nlopt}}
\end{figure*}

The parameters of ITO, including an experimental dependence of the refractive index on intensity $n(I)$ in a nonlinear regime, are taken from the literature \cite{alam-2016}. The pumping wavelength is defined by ENZ point, where a giant nonlinearity takes place. For the considered ITO parameters it is $1240$ nm. For the convenience of computation the experimental $n(I)$ dependence was fitted using the following analytic functions
\begin{equation}
\text{Re}[n(I)] = a_1\cdot\arctan(b_1\cdot{I})+c_1,
\label{eq_ren}
\end{equation}
\begin{equation}
\text{Im}[n(I)] = a_2\cdot \exp(-b_2\cdot{I})+c_2,
\label{eq_imn}
\end{equation}
where $a_1 = 0.4714$, $b_1 = 0.031$ cm$^2$/GW, $c_1 = 0.42$, $a_2 = 0.17$, $b_2 = 0.0358$ cm$^2$/GW, $c_2 = 0.25$. The fit is approximately valid for the range of intensities between $0$ and $200$ GW/cm$^2$. Thus, the refractive index of ITO is treated numerically as a general function of intensity with no usual in nonlinear optics power series decomposition \cite{boyd}. For the electrical gating efficiency calculations we modify the ITO parameters slightly to maximize the gating efficiency for a fare comparison (see the next section).

To investigate and analyze light propagation in our structures three different theoretical approaches are used. To obtain a steady state field distribution and evaluate characteristics of devices (like transmittance and phase shift) we solve numerically the wave equation for electric field $\mathbf{E}$  in frequency domain \cite{pshenichnyuk-2019}
\begin{equation}
\nabla\times\nabla\times\mathbf{E} - k_0^2\varepsilon_r\mathbf{E}=0,
\end{equation}
where $k_0=\omega/c$. Nonlinearity enters the equation through the relative permittivity $\varepsilon_r = n^2(\mathbf{r},I)$. The iterative nonlinear solver takes into account that both the field and refractive index distribution depend on each other and searches for a steady state, where both functions are mutually stabilized.

For a verification of frequency domain results we employ a time domain computational scheme, where a second order equation 
\begin{equation}
\nabla\times\nabla\times\mathbf{A} +\mu_0 \frac{\partial}{\partial t}\left[ \varepsilon_0\varepsilon_r \frac{\partial\mathbf{A}}{\partial t} \right] =0,
\end{equation}
is solved for a magnetic vector potential $\mathbf{A}$. It can be derived from Ampere's law using the following gauge conditions
\begin{equation}
\mu\mathbf{H} = \nabla\times\mathbf{A}
\end{equation}
\begin{equation}
\mathbf{E} = -\frac{\partial\mathbf{A}}{\partial{t}}
\end{equation}
Auxiliary equations are used to incorporate Drude permittivity models for ITO and Au \cite{inan}. Nonlinearity in this case enters the equations through the appropriately precalculated plasmonic frequency and Drude damping terms of ITO. For each set of parameters and each considered intensity, time evolution should be computed sufficiently far to reach a steady state, where the characteristics can be compared with the frequency domain calculations. It is the most detailed approach, but it is also time consuming. The results obtained using time domain simulations are quite close to the numbers obtained in frequency domain studies (see, for example, Fig.~\ref{fig3_nlopt}b for comparison), despite the fact that the computational scheme is quite different.

Finally, we apply a linear mode solver for the simplified analysis at the end of the next section. Intensity enters the mode equations as a parameter, that defines a constant refractive index of ITO in a whole volume using Eqs.~\ref{eq_ren}-\ref{eq_imn}. This approach neglects the influence of intensity distribution on the refractive index. But still, it allows to understand certain aspects of nonlinear behavior, like, for example, the intensity dependent hybridization between plasmonic and waveguide modes (see the next section). The drift-diffusion system of equations is implemented for the computation of charge density distribution in gated sandwiches. Please see our previous works for more details \cite{pshenichnyuk-2019}. All numerical solvers are realized in commercial software Comsol Multiphysics 5.3a. For the computations we use Lenovo P620 workstation with 64 CPU cores and 512 Gb of RAM.

\section{Results and Discussion} \label{sec_results}

In the first part of this section we are going to provide a quantitative estimation of the switching strength in nonlinear HPWG and compare it with electrical switching mechanism. In general, the possibility to switch an optical system is related to the possibility to vary its refractive index. Both the amplitude of variation and the volume fraction where changes take place should be taken into account. For instance, a variation of an average refractive index
\begin{equation}
\bar{n} = \frac{1}{V}\int\limits_{V} d\mathbf{r}\,\, n(\mathbf{r}),
\label{aveq}
\end{equation}
where the integral is taken over the volume of active material (ITO in our case), can be used for evaluation. In the electrical activation sheme the spatial distribution of the refractive index $n(\mathbf{r})$ depends on voltage $U$ and corresponding spatial distribution of electron density $n(\mathbf{r}) = n[n_e(\mathbf{r};U)]$. For the optical activation the refractive index is a function of intensity distribution, that depends on the intensity of a pump $n(\mathbf{r}) = n[I(\mathbf{r};I_p)]$. Thus, below we discuss the functions $\bar{n}(U)$ and $\bar{n}(I_p)$ depicted in Fig.~\ref{fig2_dd}c and Fig.~\ref{fig3_nlopt}a respectively.

In the electrical activation scheme the spatial distribution of electron density can be evaluated using the drift-diffuion equations. It is defined by the composition of the sandwich (Au/ITO/HfO$_2$/Si in our case) and applied voltage. Charge concentration profiles along the line perpendicular to the sandwich surface are presented in Fig.~\ref{fig2_dd}a for different voltages. The thickness of an accumulation (depletion) layer can be evaluated from these pictures. The interval where a profile variation takes place is just few nanometers thick. When the profiles are computed, Drude theory can be used to evaluate the refractive index distribution (\cite{pshenichnyuk-2019}). Corresponding quantities, both real and imaginary parts, are shown in Fig.~\ref{fig2_dd}b for different voltages. Finally, $n(\mathbf{r})$ is averaged using Eq.~\ref{aveq} to obtain $\bar{n}(U)$ plotted in Fig.~\ref{fig2_dd}c.

\begin{figure}
\centerline{\includegraphics[width=0.49\textwidth]{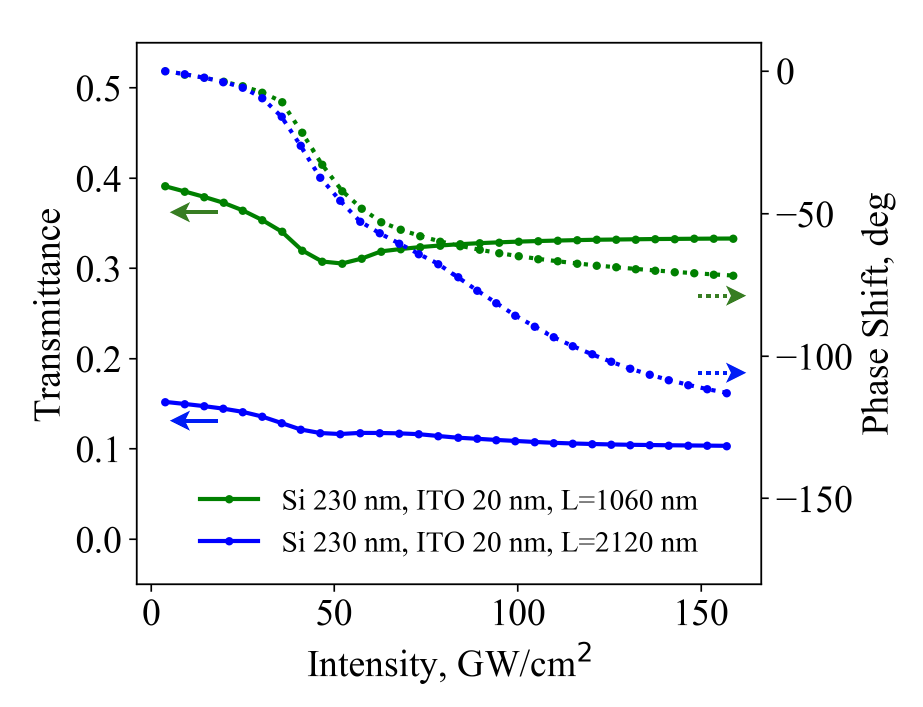}}
\caption {Transmittance (solid curves) and phase shift (dotted curves) obtained in silicon based HPWGs as a function of intensity. Green and blue curves correspond to HPWGs with single and double lengths $L$ respectively. 
\label{fig4_nlopt_si}}
\end{figure}

It is important to stress that the electrical switching strength depends on the parameters of ITO, such as the initial concentration and Drude damping $\gamma$. It also depends on the wavelength. Parameters of ITO can significantly vary in different experiments, depending on conditions \cite{kulkarni-1996}. For the optical switching evaluations (see below) we use ITO parameters from literature \cite{alam-2016}, where the nonlinear properties were investigated in ENZ regime. The charge concentration is close to $10^{21}$  cm$^{-3}$ with corresponding ENZ wavelength $1240$ nm. For the comparison with the electrical computations here we prefer to use similar ITO. On the other hand, to observe a good electrical switching one should decrease the concentration. It is necessary to provide the switching between out-of-ENZ and ENZ regimes and to make the initial ITO state more transparent. Doing evaluations, we also have to stay in a reasonable range of voltages. When $\pm 10$ V window is not sufficient to reach ENZ regime one may consider larger wavelengths where it is possible. Thus, for the evaluations below we decrease the initial concentration (that corresponds to the lower original doping level of ITO) and consider different wavelengths to maximize the gating effect for a fair comparison.

Taking this into account, we present the evaluation results in Fig.~\ref{fig2_dd}c for three different wavelengths. For the given parameters of ITO with the density decreased to $10^{20}$  cm$^{-3}$ we observe ENZ switching at approximately $2500$ nm (staying in $\pm 10$ V voltage window). Corresponding charge density profiles (Fig.~\ref{fig2_dd}a) and refractive index distributions (Fig.~\ref{fig2_dd}b) are also presented for the longest wavelength, where the effect is more evident. ENZ concentration at $2500$ nm is close to $2.6 \cdot 10^{20}$ cm$^{-3}$ (dashed horizontal line in Fig.~\ref{fig2_dd}a). It is reached at approximately $+6$ V. At two other wavelengths considered here the ENZ is not reached and the effect is more modest (see blue and green curves in Fig.~\ref{fig2_dd}c). We would like to stress that for other types of ITO with different parameters (for example, different Drude damping $\gamma$) out-of-ENZ to ENZ switching can be observed also at telecom wavelength $1550$ nm \cite{pshenichnyuk-2019}. In general, the expected strength of electrical switching is given by the amplitude of the red curve in Fig.~\ref{fig2_dd}c. Namely $0.1$ variation for the real part and $0.04$ variation for the imaginary part.

To evaluate the strength of optical switching we solve nonlinear Maxwell equations. An average intensity of the pump $I_p$ is the input parameter. Spatial distribution of the intensity at the output normally depends on the refractive index of materials. In the nonlinear formulation the refractive index also depends on intensity. During the numerical solution both distributions are mutually stabilized to reach a steady state. To check the stability of our results we used both frequency domain and time domain computations. The obtained results are quite close to each other. The comparison between frequency domain and time domain results is shown in Fig.~\ref{fig3_nlopt}b (triangular markers correspond to time domain while solid lines of the same color correspond to frequency domain). The distribution $n(\mathbf{r})$ obtained using one of the methods can then be averaged using Eq.~\ref{aveq} to compute $\bar{n}(I_p)$. It is presented in Fig.~\ref{fig3_nlopt}a.

A distribution of intensity in HPWG is related to plasmonic effects. It depends strongly on geometry and materials. The computations discussed here cover a representative set of parameters, suitable for experimental realization in our lab. For Si waveguides we consider the thickness $230$ nm, while for Si$_3$N$_4$ waveguides - $400$ nm and $500$ nm. The thickness of ITO layer is varied in $20 - 40$ nm range. The length of a modulating sandwich $L$ is different for different models (see the legends in Fig.~\ref{fig3_nlopt} and Fig.~\ref{fig4_nlopt_si}). It is defined by a plasmonic conversion length and varies between $1060$ nm and $3020$ nm. We consider both Si (Fig.~\ref{fig4_nlopt_si}) and Si$_3$N$_4$ (Fig.~\ref{fig3_nlopt}) as a materials for the waveguide. The results obtained for silicon nitride are more attractive from the point of view of applications. Slab waveguides made from Si$_3$N$_4$ have typically smaller effective indices and easier couple with SPP modes. Silicon nitride is also more transparent for short wavelengths comparing with silicon.

In all the regimes considered here, the $\bar{n}(I_p)$ dependence demonstrates a peculiar step-like behavior (Fig.~\ref{fig3_nlopt}a) useful for various switching applications. The largest variation of $\bar{n}$ is obtained for $400$ nm thick Si$_3$N$_4$ waveguide with $20$ nm thick ITO (magenta curve). Inside the selected range of pump intensities, the variation of $\bar{n}$ reaches $0.65$ for the real part and $0.16$ for the imaginary part. It is noticeably larger compared with the values obtained for the electrical switching mechanism (Fig.~\ref{fig2_dd}). Please note that the thickness of ITO in both cases is $20$ nm, but changes of the refractive index caused by the optical excitation take place in the whole volume of ITO vs few nm thick accumulation layer in the electrical case. The presence of steps is related to the regimes where SPP excitation becomes possible or vice versa, impossible (see below). It is interesting to notice that in general the position of the step moves in the direction of larger intensities together with the thickness of ITO.

\begin{figure}
\centerline{\includegraphics[width=0.5\textwidth]{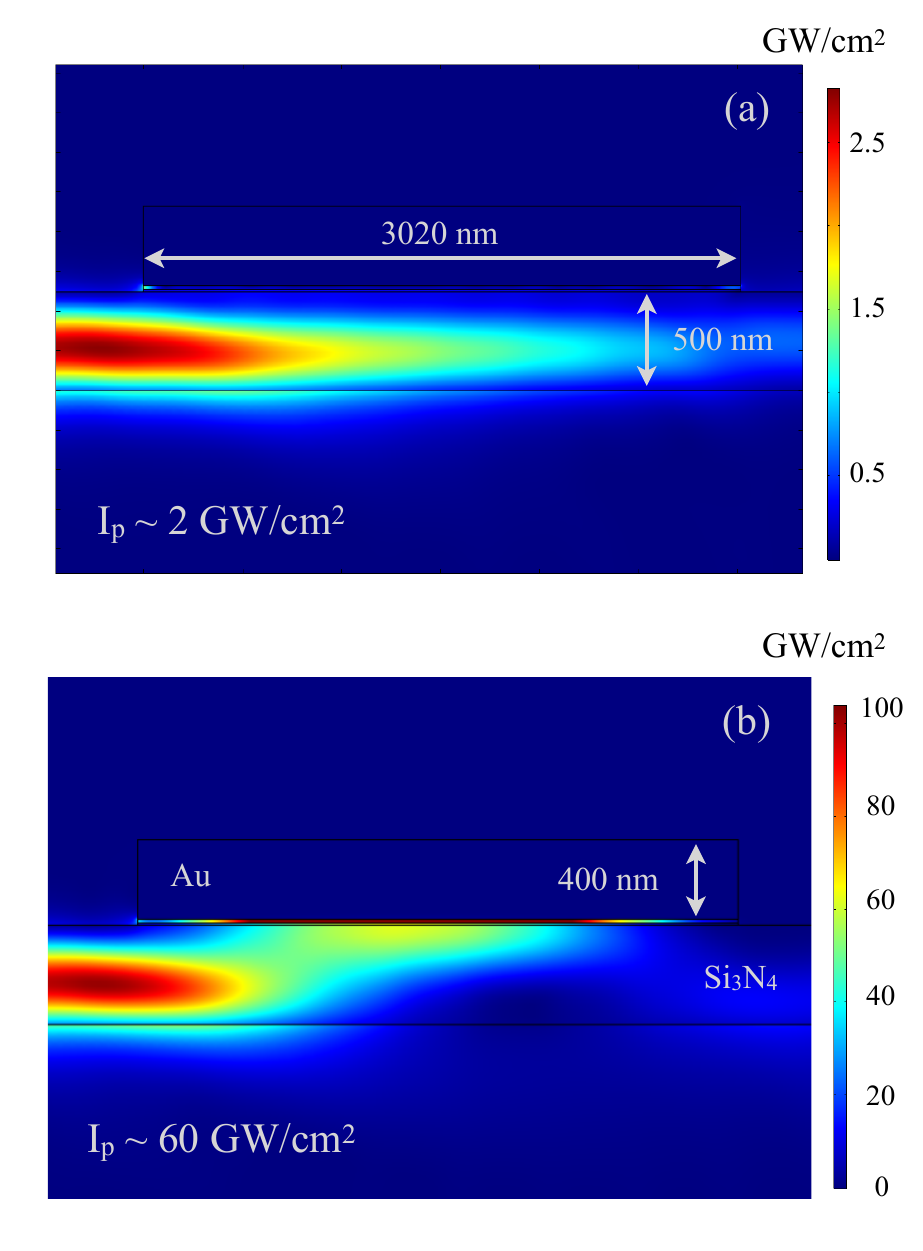}}
\caption {Power flux density distribution inside HPWG structure at small (a) and large (b) pump intensity. Note the excitation of SPP in the second case in contrast with the first case where it is forbidden. The presence of SPP provides a significant concentration of field inside the active layer of ITO (the gap between Si$_3$N$_4$ waveguide and metal).
\label{fig5_nlpl}}
\end{figure}

In the second part of this section we discuss the possibility to use nonlinear ITO based HPWG as active devices in photonic circuits. Namely amplitude modulators and phase shifters. We also analyze in more details the processes that take place inside nonlinear HPWG. ITO based electrically gated plasmonic HPWGs allow to obtain good modulating characteristics \cite{pshenichnyuk-2019,zemtsov-2023}. According to the evaluations above, optical activation mechanism in ITO potentially allows to improve characteristics of devices even further. According to Fig.~\ref{fig3_nlopt}a, the average refractive index real part variation is larger compared with the imaginary part. It suggests that nonlinear HPWG may function as effective phase shifters. The results of corresponding numerical experiments are shown in Fig.~\ref{fig3_nlopt}c for Si$_3$N$_4$ waveguides and in Fig.~\ref{fig4_nlopt_si} for Si waveguides. For the selected set of models we managed to obtain the phase shift close to $\pi$ (magenta curve in Fig.~\ref{fig3_nlopt}c). It is sufficient for various applications, including Mach-Zehnder interferometers. Note, that the length of the device is just $2550$ nm. The worst results from the point of view of phase shifting applications are demonstrated by Si waveguide (Fig.~\ref{fig4_nlopt_si}), which can be attributed to larger effective index mismatch compared to silicon nitride waveguides.

Next, we consider the relative output amplitude variation (transmittance) as a function of intensity, depicted in Fig.~\ref{fig3_nlopt}b for Si$_3$N$_4$ waveguides and in Fig.~\ref{fig4_nlopt_si} for Si waveguides. The best on/off contrast suitable for applications is demonstrated by $500$ nm thick Si$_3$N$_4$ waveguide with $40$ nm thick layer of ITO (yellow curve in Fig.~\ref{fig3_nlopt}b). Smaller but comparable contrast can be achieved in the model with slightly decreased thickness of ITO (light blue curve). Note that the position of the step shifts to the left (in the direction of smaller intensities) in this case allowing much smaller pump intensities to achieve the contrast. Again, the smallest amplitude is demonstrated by Si waveguide (Fig.~\ref{fig4_nlopt_si}). For the doubled length modulating sandwich (blue curve corresponds to two plasmon conversion lengths) the transmittance variation amplitude becomes even smaller. Interestingly, we deal here with two different types of behavior. Along with the step-like behavior we observe the situation when transmittance goes down and then goes up again (red and magenta curves in Fig.~\ref{fig3_nlopt}b). It takes place for thinner layers of ITO and is attributed to a tricky interplay between various modes (see below in this section). It is important to note that the best amplitude modulation regime (yellow curve in Fig.~\ref{fig3_nlopt}) and the best phase shifting regime (magenta curve in Fig.~\ref{fig3_nlopt}) correspond to different geometries.

To illustrate the behavior of light inside a nonlinear HPWG we plot the power flux density distribution at high and low pump intensities in Fig.~\ref{fig5_nlpl}. The plots correspond to $500$ nm thick Si$_3$N$_4$ waveguide model (yellow curve in Fig.~\ref{fig3_nlopt}). It is clear from the picture that SPP excitation does not take place at small intensities. Moreover, the field is slightly pushed down out of the sandwich (Fig.~\ref{fig3_nlopt}a). On the other hand, at high intensities we observe strong SPP excitation. The field becomes transformed into SPP where it loses a certain fraction of power and then returns into a waveguide. In the second scenario, due to plasmonic amplification and subwavelength confinement inside the sandwich, interaction of ITO with the field becomes more intensive. We stress out that the active material in nonlinear HPWG is originally prepared in ENZ state and, consequently, it is highly absorptive. Despite the fact that the imaginary part of ITO refractive index goes down with intensity (Fig.~\ref{fig3_nlopt}a), the overall effect for the particular $500$ nm thick (yellow)  model is the step like decrease of transmittance (Fig.~\ref{fig3_nlopt}b). In general, the interplay between these two factors may cause other types of $T(I)$ dependence, like the one represented by the magenta curve in Fig.~\ref{fig3_nlopt}b.

Note that a thick enough  layer of nontransparent ITO provides 'screening' for SPP carrying golden surface above, making the plasmon excitation impossible. But this constraint can be suppressed at higher intensity, where ITO becomes more transparent. Thus, nonlinear HPWG can realize a mechanism where switching occurs between the presence and absence of surface plasmons. It is different to compare with the electrical gating mechanism, where a thin accumulation layer may be used to vary a degree of absorption and probably conversion length of SPP but may not completely suppress plasmon excitation \cite{pshenichnyuk-2021}. Since the appearance of SPP is related to increased losses, it would be nicer to have SPP at low intensities and suppress it at high intensities for practical amplitude modulation tasks. In this case two factors mentioned above, namely the low transparency at low intensities and the presence of SPP, would sum up and provide larger modulation contrast. Searching for such a regime is an interesting task for future works.

\begin{figure}
\centerline{\includegraphics[width=0.5\textwidth]{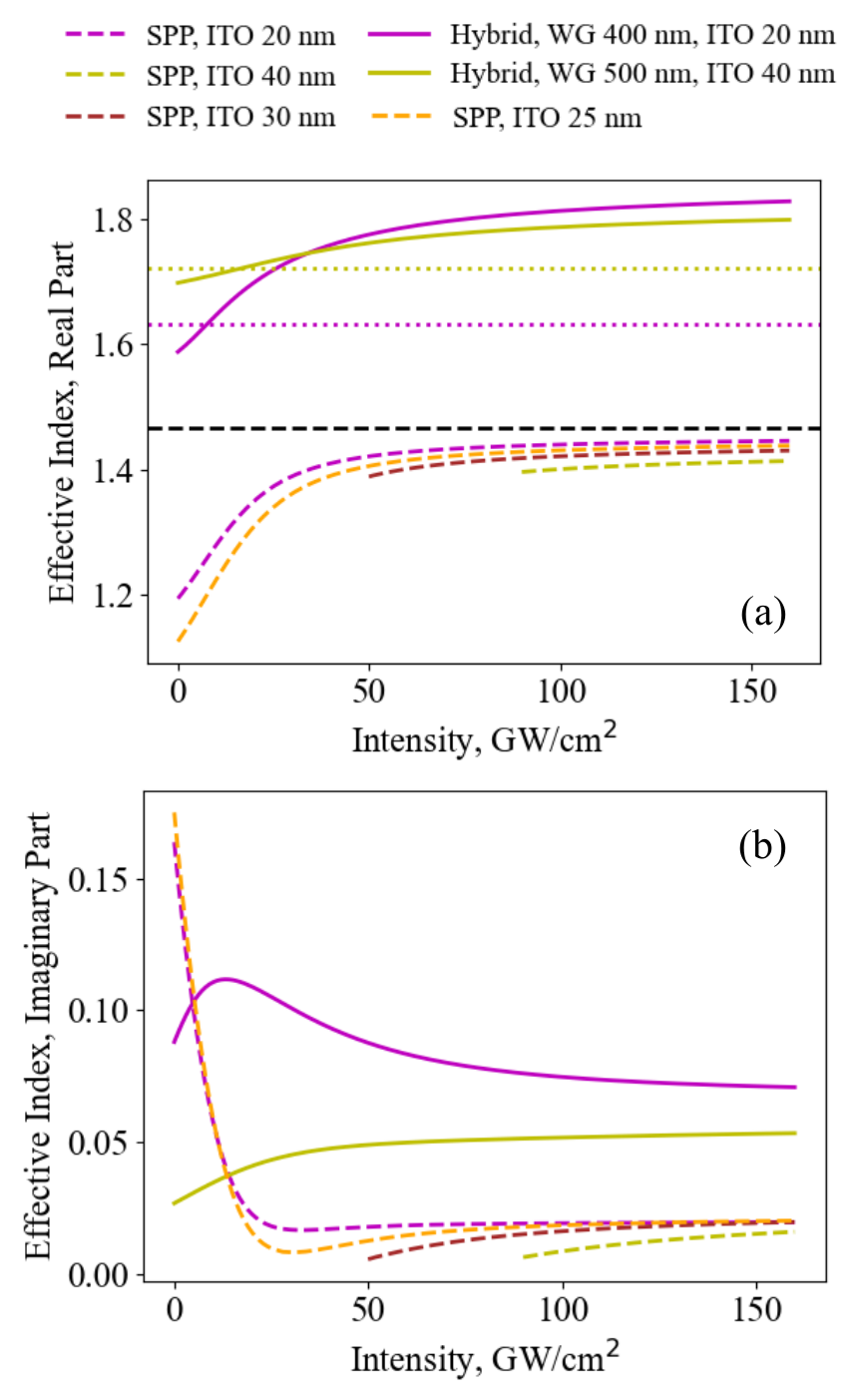}}
\caption {Hybridization between SPP (dashed curves) and waveguide modes (dotted curves) forms hybrid modes (solid lines). Real (a) and imaginary (b) parts of the effective indices are demonstrated. Yellow and magenta color codes coincide with corresponding notations in Fig.~\ref{fig3_nlopt}. Black dashed horizontal line in (a) represents the effective index of SPP in the absence of ITO. 
\label{fig6_linan}}
\end{figure}

To analyze plasmon excitation conditions and understand better the form of $T(I)$ curves we perform a linear mode analysis for selected structures. In these simplified calculations we introduce 'intensity' as a parameter that defines a homogeneous refractive index for a whole volume of ITO. It is used as an input parameter for linear mode equations. Mutual dynamical influence of light intensity distribution and related inhomogeneous refractive index is not considered in this approach. Mode analysis also assumes a homogeneous structure in the direction of propagation thus excluding waveguide-sandwich-waveguide transition process from the consideration. The computed effective indices of structures are plotted in Fig.~\ref{fig6_linan} as a function of intensity parameter. Please mind the difference between horizontal axis in this figure and the axis in Fig.~\ref{fig3_nlopt} where 'intensity' corresponds to the mean pump intensity $I_p$ used in nonlinear models.

Yellow and magenta curves are chosen from the original set of models (Fig.~\ref{fig3_nlopt}) to perform the mode analysis since they represent two qualitatively different types of behavior (described above).  Along with the hybrid modes of full structures (solid curves) we consider pure surface plasmons when the Si$_3$N$_4$ slabs are removed (dashed curves). In addition to the original models, we also consider two more thicknesses of ITO, namely $25$ nm (orange dashed curve) and $30$ nm (brown dashed curve). The effective index of pure SPP with completely removed ITO is shown by horizontal dashed black line in Fig.~\ref{fig6_linan}a. Clearly, it is intensity independent in our model. At the same figure we also add the effective indices of $400$ nm and $500$ nm silicon nitride slabs (fundamental TM modes) plotted with dotted horizontal magenta and yellow curves respectively.

One important conclusion that follows from the results shown in Fig.~\ref{fig6_linan} is the presence of a minimal thickness of ENZ ITO that is required to screen the golden surface and prevent the excitation of SPP. For $20$ nm and $25$ nm ITO layer thicknesses (magenta and orange dashed curves) our numerical algorithm recognizes SPP modes. For larger thicknesses, starting from $30$ nm (brown dashed curve), we cannot detect any modes at small intensities. On the other hand, when we increase the intensity, SPP modes can be detected. Moreover, in accordance with the results in Fig.~\ref{fig3_nlopt}a, there exists an intensity threshold that becomes larger along with the thickness. Note that when we increase the intensity (and transparency of ITO) the screening effect becomes weaker and the effective index of corresponding SPP approaches the effective index of pure SPP (Fig.~\ref{fig6_linan}a, horizontal black dashed line).

When we include a waveguide into consideration, the SPP modes (dashed lines) hybridize with waveguide modes (horizontal dotted lines). Note that the vertical distance between the lines is not large (Fig.~\ref{fig6_linan}). For Si waveguides the distance is larger because of the large Si refractive index and the hybridization is less effective. The resulting hybrid modes are shown using solid magenta and yellow lines. The imaginary part of effective index (Fig.~\ref{fig6_linan}b) is responsible for losses. For the magenta curve the losses originally grow with intensity but then start to decrease after some point. For yellow curve losses always monotonically grow with intensity. It is in correspondence with the nonlinear results in Fig.~\ref{fig3_nlopt}b (magenta and yellow curves). Thus, two types of behavior mentioned above can be identified in terms of the simplified linear model as different types of hybridization between ITO screened SPP and waveguide modes. At the same time, the peculiar step-like switching depicted in Fig.~\ref{fig3_nlopt}a is a nonlinear effect that cannot be analyzed using linearized models.

\section{Conclusion} \label{sec_conclusion}

The possibility to realize an all-optical switching scheme in a nonlinear HPWG and use it as a basis for active devices in integrated photonics is investigated. The mechanism is related to the giant optical nonlinearity of ITO in ENZ regime. Three different theoretical approaches are used for the analysis: frequency domain and time domain nonlinear Maxwell equations, and the linear mode solver. A comparison between all optical nonlinear HPWG switching efficiency and electrical gating switching scheme is performed. It is shown that the optical scheme may potentially provide stronger switching. The reason is that the optical pumping may influence a larger volume of ITO layer, while the gating can only be used to manipulate a thin accumulation layer at ITO/insulator boundary. The basic possibility to use nonlinear HPWG for the manipulation of phase is demonstrated, $\pi$ phase shift is achieved for just $2.5$ $\mu$m long device. Amplitude modulation is also demonstrated and discussed. Valuable for switching applications step-like dependence of the transmittance on intensity is shown. The possibility to use the intensity of light to allow and prohibit SPP excitation is demonstrated and analyzed. Subsequent implementation and improvement of the switching scheme for particular applications is a subject of future works.

\bibliography{paper_nonlin}

\end{document}